\documentclass[lettersize,journal]{IEEEtran}
\usepackage{amsmath,amssymb,amsfonts}
\usepackage{algorithmic}
\usepackage{algorithm}
\usepackage{array}
\usepackage[caption=false,font=normalsize,labelfont=sf,textfont=sf]{subfig}
\usepackage{textcomp}
\usepackage{stfloats}
\usepackage{url}
\usepackage{verbatim}
\usepackage{graphicx}
\usepackage{cite}
\usepackage{color}
\usepackage{dutchcal}
\usepackage{cancel}
\usepackage{gensymb}
\usepackage{blindtext}
\usepackage{booktabs}
\usepackage{lscape}
\usepackage{multirow}
\usepackage{hyperref}

\begin{document}

\title{An Optimal Task Planning and Agent-aware Allocation Algorithm in Collaborative Tasks Combining with PDDL and POPF}

\author{Qiguang Chen,~\IEEEmembership{Student Member,~IEEE,} and Ya-Jun Pan,~\IEEEmembership{Senior Member,~IEEE}
        % <-this % stops a space
\thanks{Thanks to the Natural Sciences and Engineering Reserach Council of Canada (NSERC), and the Government of Nova Scotia for supporting this work. The authors are with the Department of Mechanical Engineering, Dalhousie University, Halifax NS B3H 4R2, Canada (e-mail: qg234631@dal.ca; yajun.pan@dal.ca).}}

% Remember, if you use this you must call \IEEEpubidadjcol in the second
% column for its text to clear the IEEEpubid mark.

\maketitle

\begin{abstract}
In Industry 4.0, it proposes the integration of artificial intelligence (AI) into manufacturing and other industries to create smart collaborative systems which enhance efficiency. The aim of this paper is to develop a flexible and adaptive framework to generate optimal plans for collaborative robots and human workers to replace rigid, hard-coded production line plans in industrial scenarios. This will be achieved by integrating the Planning Domain Definition Language (PDDL), Partial Order Planning Forwards (POPF) task planner, and a task allocation algorithm. The task allocation algorithm proposed in this paper generates a cost function for general actions in the industrial scenario, such as \textbf{PICK}, \textbf{PLACE}, and \textbf{MOVE}, by considering practical factors such as feasibility, reachability, safety, and cooperation level for both robots and human agents. The actions and costs will then be translated into a language understandable by the planning system using PDDL and fed into POPF solver to generate an optimal action plan. In the end, experiments are conducted where assembly tasks are executed by a collaborative system with two manipulators and a human worker to test the feasibility of the theory proposed in this paper.
\end{abstract}


\section{Introduction}
\IEEEPARstart{R}{ecently}, the Fourth Industrial Revolution or simply “Industry 4.0” (I4.0), has been transforming the way  industries manufacture their products and deliver their services\cite{b1}. It refers to the integration of advanced digital technologies such as artificial intelligence, Internet of Things (IoT), and data analytics, introducing it into manufacturing processes to create smarter, more connected, and efficient industrial systems\cite{b2, b3}. The traditional requirement of producing thousands of identical products on a production line has evolved into producing thousands of products with similar functions and different appearance tailored to fulfill customer demands\cite{b4}.

Task planning in industry refers to the process of organizing and scheduling activities within manufacturing processes \cite{b5}. It can improve the rigid traditional production lines by helping machines perform actions needed for completing similar tasks with various parameters and flexible orders to meet diverse customer needs efficiently \cite{b6, b7, b8}. The article focuses on two main aspects of task planning: the language used and the task planner employed. We select PDDL and POPF for our approach. The PDDL stands as one of the most classic task planning languages, it serves the purpose of translating production line actions and plans into a code format that is readily comprehensive for computers and machines \cite{b9, b10}. PDDL serves as the artificial intelligence environment for automated planning of production processes integrated with Industry 4.0 standards, to establish a foundational model for production systems \cite{b11}. \cite{b12} uses PDDL as the bridge between automation and AI planning to enable a realistic automated planning and scheduling in discrete manufacturing.

Partial Order Planning Forwards (POPF) is a forward-chaining temporal planner that integrates partial-order planning concepts\cite{b13}. It is widely used in the area of robot system \cite{b14, b15}. One major advantage of POPF is its capability to plan actions with cost considerations. This enables the generation of plans that not only optimize logical order but also minimize action costs. For this reason, POPF emerges as a suitable task planner in scenarios where the appropriate agents are required for assigning specific tasks.

As one of the most pivotal type of robotic agents in smart factories, robotic manipulators play a central role in Industry 4.0. Increasing number of articles are combining task planning algorithms with manipulators to create intelligent work systems \cite{b16}-\cite{b18}. In \cite{b19}, a bi-level lazy search is guided by PDDLStream to make improvements in planning efficiency and success rates over existing solvers with a 7-DoF manipulator. \cite{b20} proposes using LEMMA with PDDL to study vision-based language-conditioned multi-robot collaboration with several manipulators. In our approach, we rely on task planning algorithms as the primary overseer for assembly tasks. It will generate an optimal plan and direct both robots and humans to carry out their assigned tasks.

With the advancements in mobile robots and manipulators, they are now capable of fulfilling roles traditionally performed by humans more effectively than ever before. Determining the most suitable agent for a specific task has emerged as a hot topic of discussion recently.

The debate over which agent is the optimal choice for specific tasks remains highly relevant in the era of Industry 4.0. The task allocation algorithm for calculating specific costs incurred by an agent to complete a certain task serves as a critical method to identify the most suitable agents for the job. \cite{b17} suggests an approach for task allocation, taking into account the real capabilities of both humans and robots, aiming to enhance the quality of work. \cite{b21} designed a task allocation algorithm involving integrating task complexity, agent dexterity, and agent effort to ascertain the most suitable agent, manipulator or human, for each step in a manufacturing assembly task. In our approach, the task allocation algorithm is built around collaborative scenarios, wherein each agent completes specific tasks. This algorithm is then fed into the POPF task planner, which not only generates a task plan with an optimal logical sequence but also guarantees that each action is delegated to the most suitable agent for execution. To the author’s knowledge, PDDL and POPF has not been integrated with task allocation algorithms and applied to a real life scenario with multiple manipulators and human.

The aim of this article is to generate an optimal plan for collaborative robots to replaces rigid, hard-coded production line plans with a more flexible and adaptive system in collaborative industrial scenarios. Three main innovations are listed to meet the aim of this article:

1. An action library comprising general actions for both humans and manipulators in collaborative industrial scenarios is established and transformed into PDDL to make sure the actions library can be understood by computer and machine.

2. The cost function of each action is designed based on the feasibility, reachability, and safety for each agent, and cooperation level when dealing with cooperative actions. The task allocation algorithm is inputted to the POPF task planner to not only create a task plan with an optimal logical order but also ensures that each action is assigned to the most suitable agent for execution.

3. For demonstrations, the integrated algorithm is applied to a real-time collaborative environment, including two manipulators and a human worker, to test its feasibility.

The rest of this article is organized as follows. In Section II, the problem formulation and motivation of the work are illustrated. The methodology of using the proposed task planning algorithm is introduced in Section III. Section IV presents the design of task allocation algorithm. Experimental results are carried out in Section V. Finally, Section VI concludes this article. 

\section{Problem Formulation}
In the automation era, factories primarily rely on rigid production line processes to mass-produce large amount of identical products. As a critical type of industrial robot agents, robot manipulators can execute tasks according to pre-programmed instructions such as moving trajectory, and pick-and-place actions, coded by human programmers. Additionally, humans participate in task completion by operating robots or working alongside them, though robots can also operate independently. To align with the smart factory concept and the new Human-Robot Interaction (HRI) model proposed by Industry 4.0, new features are being pursued by using task planning algorithm, and task allocation algorithm. Instead of completing tasks through rigid programming and relying on human workers to cooperate based on their experience, a commander-like task planning algorithm is needed.  This algorithm would have the capability of generating a comprehensive plan, ensuring the right agent performs the right task at the right time. Such an approach would effectively manage the completion of complex logistic tasks, which involve multiple subtasks to be executed in a specific order. An example could be found in \cite{b18} where a flexible planning algorithm based on PDDL is developed to command the robot manipulator to finish similar tasks but with different configuration such as making cocktails following different recipes and customer's instructions by altering the order and/or modifying the parameters of robot manipulator's actions.

\section{Task Planning Algorithm}

This section introduces using task planning algorithms (PDDL and POPF) to organize and execute tasks more efficiently. The innovation of this section is demonstrating a structure to show how to integrate task allocation algorithms with task planning algorithms.

\begin{figure}[h!]
\centering
\includegraphics[width=8.5cm]{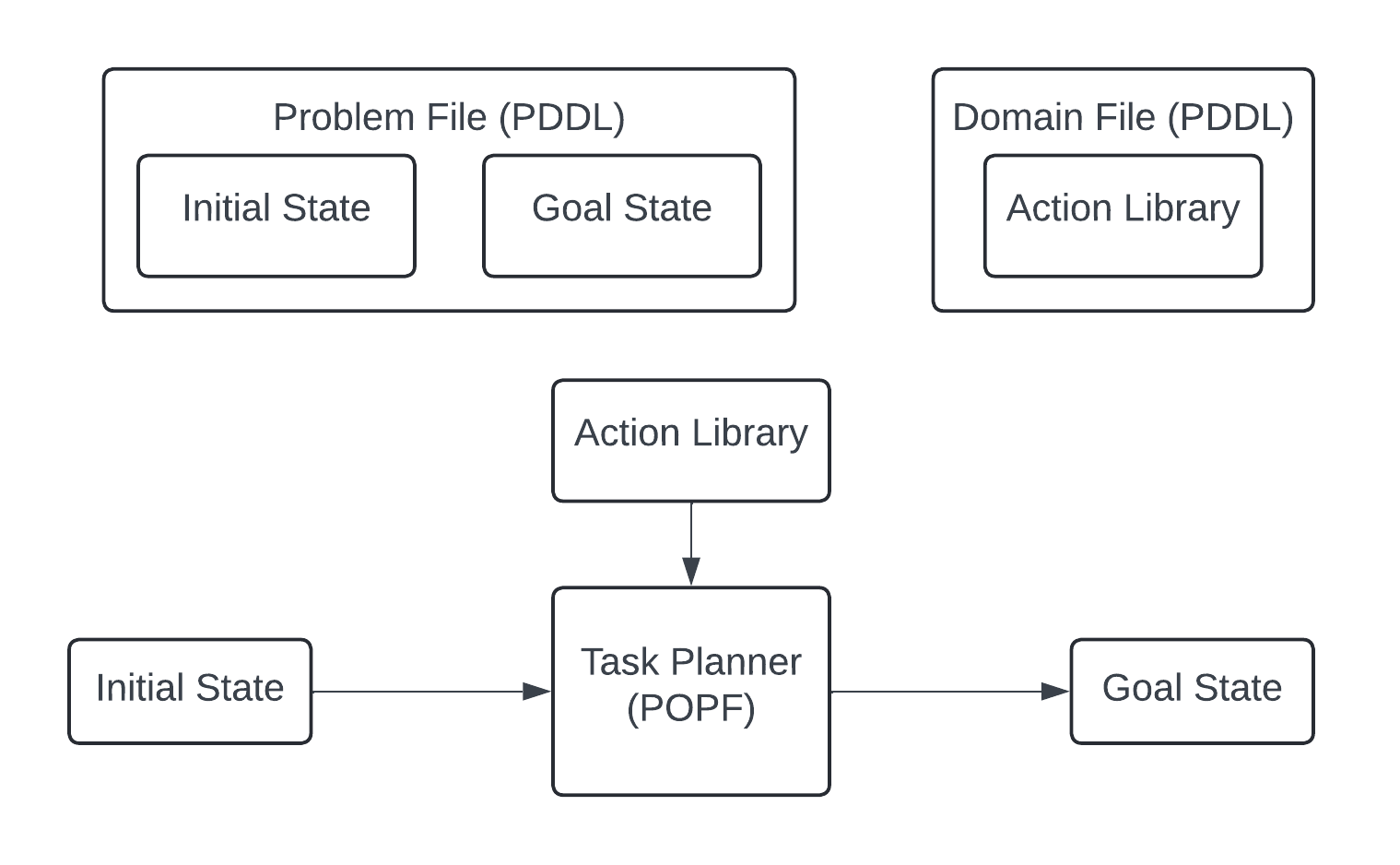}
\caption{A block diagram of task planning algorithm}
\label{A Block Diagram of Task Planning Algorithm}
\end{figure}

As shown in Fig. \ref{A Block Diagram of Task Planning Algorithm},  a typical use of PDDL involves two files: a domain file and a problem file. A basic domain file can be used to define three objects: parameters, $\mathcal{O}$, predicates, $\mathcal{P}({O_1}, {O_2}, \ldots, {O_n})$, and actions, $\mathcal{A}({O_1}, {O_2}, \ldots, {O_n})$. The parameter simply indicates the existence of this parameter within the domain file. Predicates is properties or relationships that can hold true or false in the domain, which can be used for the preconditions and effect for actions. Actions can be taken in the domain, along with their preconditions and effects. Precondition and effect belong to two  adjacent states of system, while apply an action at the ${n^{th}}$ state, $\mathcal{S_n}$, can transit it to the next state, $\mathcal{S_{n+1}}$.

% objective

The problem file introduces an initial state, $\mathcal{S}_{initial}$, and goal state, $\mathcal{S}_{goal}$, of the whole system. The role of the task planner is to achieve  the $\mathcal{S}_{goal}$ from $\mathcal{S}_{initial}$ by performing  a sequence of actions programmed in the action library as shown in Fig. \ref{A Block Diagram of Task Planning Algorithm}. 

\begin{figure}[h!]
\centering
\includegraphics[width=8cm]{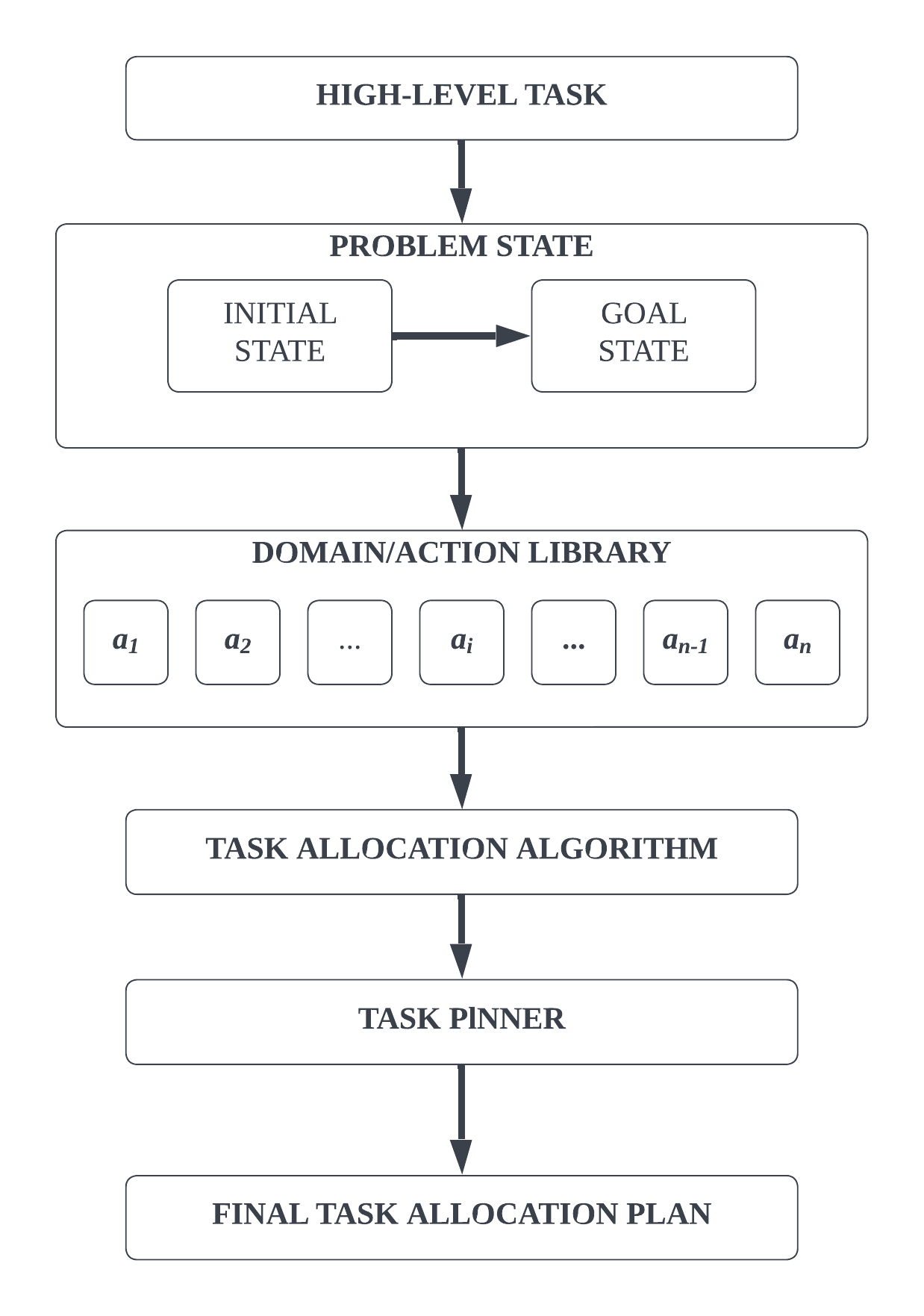}
\caption{Illustration of the original Fitts list}
\label{Illustration of the original Fitts list}
\end{figure}

Fig. \ref{Illustration of the original Fitts list} shows an approach to integrate a task allocation algorithm with a task planning algorithm. The first step is to identify the initial state and goal state of a high-level task. For example, in a smart factory, the initial state includes initial information of all the raw materials, agents, equipment and their respective locations, while the goal state comprises various products ordered by customers. To achieve the goal state from the initial state, an action library is built using PDDL. This library includes agents' capable  actions such as \textbf{PICK}, \textbf{PLACE}, and \textbf{MOVE}, which are used to achieve the goal state. Following this, a task allocation algorithm is employed to determine the cost of each specific action with different parameters. For instance, a human worker cannot lift tons of steel, so the cost for the action LIFT (Human, Steel) will be set to infinity in this case, ensuring that the task planner never assigns this task to a human agent. Next, the cost will be assigned to each action, and the updated PDDL domain will be inputted into the POPF task planner. Ultimately, the task planner will generate an optimal final task allocation plan.

\section{Task Allocation Algorithm}
\label{task_allocation}

This section aims to calculate the cost of various actions based on different parameters, such as the agent, $A$ and involved part, $P$, and location $Pos$. The innovation of this section is that an overall cost of an action is determined by factors including action feasibility, agent's reachability, agent's safety, and the level of cooperation required.  In \cite{b22}, general actions for collaborative robots in industrial environments are outlined. To obtain appropriate costs of each action under different parameters for the task planning algorithm, this article simplifies actions in \cite{b22} to \textbf{PICK}, \textbf{PLACE}, \textbf{MOVE}, and \textbf{COOPERATE}, where \textbf{COOPERATE} is for two agents doing one action together.

\subsection{Feasibility}
\label{feasibility}
In this section, the term feasibility refers to an agent’s capability of handling a specific action. For example, a human worker cannot lift and move a part heavier than the worker’s capacity, and an agent cannot reach something outside its vision field. Three sections are considered for feasibility, which are agent strength, system`s information level, and agent flexibility.

\subsubsection{Agent Strength}
\label{agent_strength}
Agent’s strength has been a major limitation of human agents to perform tasks involving high-weight objects such as heavy metal parts. To avoid physical damage on workers (mainly musculoskeletal injuries), various studies on worker’s weight capacity have been conducted. \cite{b23} introduces a general formula to determine worker lifting capacity based on various worker conditions including age, body mass index (BMI), and specific muscular conditions. \cite{b24} proposes a strength criterion to evaluate agent’s strength based on part weight and worker strength limit or robot payload.
Based on that, the expression of  strength criterion of a certain action where a part with certain weight is involved for an agent can be determined as following:
\begin{equation}
{S_a}(A_i, P_j)= 1-\frac{W_P}{S_L},
\label{strength_equation}
\end{equation}
where $S_a$ is the strength criterion of a certain action for an agent $A_i$ (unitless). $W_P$ is the weight of the involved part $P_j$ with unit of kilograms. $S_L$ is the agent's strength limit, or robot manipulator's indicated payload in kilograms. This criterion represents how close is the part weight to the agent's strength limit. 

When the strength criterion is determined for a specification, following algorithm can determine the weight-loading cost of assigning a certain action to a certain agent: 

\begin{equation}
F_S(A_i, a_j, P_k)=\left\{
\begin{aligned}
&& 0, &&  S_{A} > 0 \\
&& \infty,  && S_{A} < 0, \\
\end{aligned}
\right.
\nonumber
\end{equation}
where ${F_S}$ is the weight-loading section of feasibility cost, ${S_A}$ is the strength criterion based on the difference between the human worker strength limit and robot manipulator payload. ${F_S}$ will be set to 0 when an agent can meet the strength requirement of a certain task. If the agent does not satisfy the strength requirement, ${F_S}$ will be set to infinity.

\subsubsection{System`s Information Level}
\label{system_information_level}
With the lack of knowledge of the environment and global state of the system, robot manipulators are not capable of handling interactive tasks. For example, a robot manipulator cannot perform a pick-and-place task on a certain object without knowing the coordinate of the object and/or the target location. It is necessary to provide environmental information to the robot with sensors or human remote control/demonstration. And then, robot manipulators can perform path and trajectory planning with advanced algorithms or can simply replay the trajectory of human demonstration with knowledge of corresponding environment information to finish certain tasks. The criterion of system information level is set to be a binary criterion reflecting if robot manipulators are receiving necessary information to complete the desired action, such as vision, desired interactive force, or grasping force. 
To obtain the criterion, the following sets are defined:
\begin{itemize}
    \item $I_r (A_i,a_j )$: a set of information required by robot manipulator $A_i$ to complete specific action  $a_j$; 
    \item $I_k (A_i )$: a set of robot manipulator $A_i$’s known information.
\end{itemize}
Therefore, the information level criterion for robot agent $A_i$ to complete action $a_j$, denoted by $F_I (A_i,a_j)$ as:
\begin{equation}
F_I(A_i, a_j)=\left\{
\begin{aligned}
0, & &  if  && I_r \in I_k, \\
\infty, & & if  && I_r \notin I_k. \\
\end{aligned}
\right.
\end{equation}
Simply, if the robot agent has already gathered all information required to complete the action, the action will be assigned to the robot agent; when some information is missing, the action will be assigned to other agents or execute prerequisite actions to gather required information for that action. Because an action series will be provided by the task planning algorithm, during the process of agents performing tasks, the set $I_k (A_i)$ will be updating while other tasks are being executed to unlock the actions witch require more environment information.

\subsubsection{Agent Flexibility}
\label{agent_vlexibility}
The purpose of this section is to design a cost function to determine whether a certain agent should perform a task based on whether the task object is within the agent's range. The cost in this section is set to zero for a human agent due to human's capability of moving anywhere within the workspace. However, a robot manipulator has its limitations: it can't execute actions outside of its range. With actions considered in this paper, actions \textbf{PICK}, \textbf{PLACE}, and \textbf{COOPERATION} can be finished in a certain range of coordinates, and the coordinates of the farthest point (from the base of the involved robot manipulator) of the action \textbf{MOVE} is considered to be the one used to determine whether a manipulator agent should be assigned the task.

The flexibility criterion for a robot manipulator to perform an action can be derived as following:
\begin{equation}
\label{range_equation_new}
F_R=\left\{
\begin{aligned}
0, & &  if  &&  {x_P \in X_R}  \And {y_P \in Y_R}  \And {z_P \in Z_R}\\
\infty, & &  if  &&  {x_P \notin  X_R}  \lor {y_P \notin Y_R}  \lor {z_P \notin Z_R}\\
\end{aligned}
\nonumber\right.
\end{equation}
where $x_P$, $y_P$, and $z_P$ are x-, y-, and z-coordinates of the involved part; $X_R$, $Y_R$, and $Z_R$ are pre-determined robot manipulator range in a form of sets of coordinates within the robot manipulator range. The aim of this function is to avoid assigning actions involving points out of robot manipulator's available range.

\subsection{Reachability}

As a fundamental property of an agent, the agent reachability does not only include whether the target is within the agent’s range of reach. In an industrial environment, the agent’s posture while performing action is also an important consideration. For a robot agent, although the object is within the indicated range, a too-near or too-far position will lead to an awkward grasping posture with extreme joint angles and low controllability of arm segments, causing risk of failure and unnecessary damages. A lot of advance algorithms are presented in this area. However our approach is mainly focusing on providing a feasible cost function to the task planner for task assigning, so a simpler cost function will be designed for this section.

Lots of researchers have proposed different approaches to solve robot’s reachability within its indicated range to avoid awkward and unstable grasping postures, one of the most general method is using a capability map \cite{b25, b26, b27, b28}. 

\begin{figure}[h!]
\centering
\includegraphics[width=6cm]{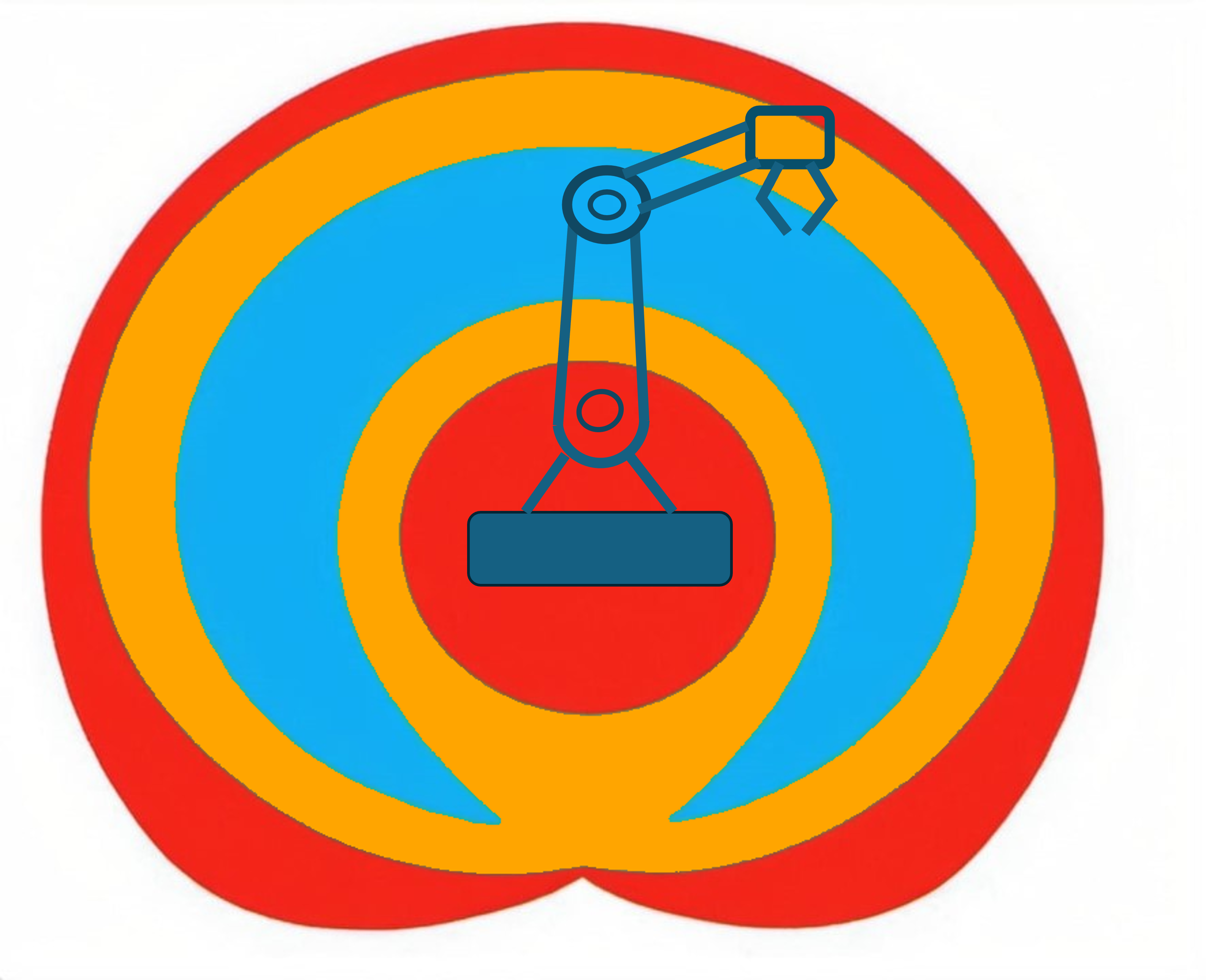}
\caption{Capability cost of a typical 7-DOF robot arm. The colors indicate the suitability of allowing a manipulator to operate in certain areas: blue represents the most suitable region, orange represents a suitable region, and red represents an unsuitable region.}
\label{image_capability_map}
\end{figure}

Based on Zacharias' capability map theory, when the robot manipulator's workspace is discretized into spheres with same dimension, the reachability index $D$ can be expressed as: 

%不要红绿配 green to something
%红 to 暗红

\begin{equation}
\begin{aligned}
D(x_P,y_P,z_P) = \dfrac{R}{N} * 100,
\end{aligned}
\label{reachability_index}
\end{equation}
where $N$ is number of randomly-sampled points on the surface of a sphere, $R$ is the number of valid inverse kinematics solutions on the surface of one sphere. The reachability section of the task allocation algorithm is derived as Eq.(\ref{reachability_index}):

\begin{equation}
\label{robot_reachability_equation}
R_R(A_i, a_j, Pos_P)=\left\{
\begin{aligned}
0, & &  if  && D > 60,\\
1-\frac{R}{N}, & & if && 20 < D \leq 60,\\
\frac{N}{R}, && if && D \leq 20.\\
\end{aligned}
\nonumber\right.
\end{equation}

Based on Eq.(\ref{reachability_index}), a robot  manipulator's workspace can be divided into three types of regions with different reachabilities:
\begin{itemize}
    \item \textbf{Most suitable region}: points with reachability index greater than 60\%. Points within the region can be approached from almost all directions, no extra cost is added to certain action.
    \item \textbf{Suitable region}: points with reachability index between 20\% to 60\%. Points within the region can be approached from most of directions and it is expected that most of robot manipulator actions will be performed in this region. A low extra cost will be added.
    \item \textbf{Unsuitable region}: points with reachability index less than 20\%, which indicates only few approachable directions are available and unstable grasps are more likely to happen since grasp postures are limited. Large cost will be added to avoid frequently performing actions in this region.
\end{itemize}

Fig. \ref{image_capability_map} shows a capability cost map of a typical 7-DOF robotic arm which is generated based on Eq (\ref{reachability_index}). %The colors indicate the suitability of allowing a manipulator to operate in certain areas: blue represents the most suitable region, orange represents a suitable region, and red represents an unsuitable region.

\subsection{Safety}

As introduced in the last section, a manipulator without the support of external sensors has no capacity of responding to the environment. Therefore, it could be dangerous by allowing an agent to enter a working manipulator's workspace. The purpose of this section is to design a cost function to reduce the preference of letting an agent work in the workspace of a manipulator. 

% One possible approach to include the safety consideration in the task allocation algorithm is to develop a risk matrix for each action and each agent. Considering the human-robot collaboration environment introduced before, the risk criterion for an agent completing a certain type of action, denoted by $C_K(A_i,a_j)$ is listed in Table. \ref{risk_matrix}.
One possible approach to include the safety consideration in the task allocation algorithm is to derive an extra safety criterion integrated into the overall cost function. Considering the human-robot collaboration environment introduced before, the unitless risk criterion for an agent completing a certain type of action, denoted by $C_K(A_i,a_j)$.
The criterion represents the level of risk of accident. It is normally considered that the risk is higher for robots due to their lower mobility and flexibility comparing to human agents. And the criterion for action \textbf{MOVE} is normally higher than that for \textbf{PICK/PLACE} since the action \textbf{MOVE} is considered to have a higher risk of intersecting with other agents.

Since human workers have higher mobility in the workspace, it is necessary to avoid the risk of getting intersection with a working robot manipulator by avoiding assigning actions involving worker's body passing through a robot manipulator's working range. An intersection coefficient only for human workers is derived as following:

\begin{equation}
    \label{intersection_equation}
    \centering
    C_I(T_A) = \left\{
    \begin{aligned}
        0, && if && T_W \cap R_{robot} = \emptyset,\\
        K_{C,i}, && if && T_W \cap R_{robot} \neq \emptyset,\\
    \end{aligned}
    \right.
\end{equation}
where $T_W$ is the planned trajectory for a human worker which is a set containing a series of coordinates. If the set of coordinates intersects with the robot manipulator's range, $R_{robot}$, an extra coefficient will be added to the safety cost for the human worker performing the action and planner will attempt to find another plan.  $K_{C,i}$ is the gain used to control the additional cost value for the situation where the human worker has to move their body and enter manipulator`s workspace. The upper limit of $C_I$ is not set to infinity since although there is a risk of intersection, human should also be able to avoid it for most of the time.

Another important consideration in the aspect of safety is robot manipulator's reachability. As introduced previously, lower reachability cost suggests increased manipulators' level of activity in that area, collisions with other agents more likely to happen. Therefore, it is necessary to introduce the reachability cost from previous section into safety criterion. The expression of safety cost of a certain action being performed by an agent, denoted as $C_S(A_i,a_j)$:
\begin{equation}
\label{safety_equation}
\centering
C_S(A_i, a_j) = \left\{
\begin{aligned}
(1+C_K)*\frac{D}{100}, && if && A_i \in A_{robot},\\
(C_I + C_K)*\frac{D}{100}, && if && A_i \in A_{human},\\
\end{aligned}
\nonumber\right.
\end{equation}
where $D(x_P,y_P, z_P)$ is the reachability index as in Eq.(\ref{reachability_index}).

\subsection{Cooperation Level}

There have been lots of researches introducing cooperation between multiple robots or human in one team to finish one action. \cite{b29} introduces how to cooperatively manipulate a single object with multiple robot manipulators by integrating the admittance control, non-singular terminal sliding mode (NTSM) control, and load distribution algorithms. \cite{b30} proposes a method that allows the robot to adapt its physical behaviour to the human fatigue in human-robot co-manipulation tasks. Although more and more advanced algorithms are applied in this area, most of them stopped at the stage of simulation, or external sensors are required to provide real-time position and force feedback information of partners to robot agents, which cause large limitation to such actions. The purpose of the method in this section is to reduce the preference of assigning an action to more than one agent, if it is feasible for only one agent.

Three combinations are made between robot agent and human agent: Human-Human Interface, Human-Robot Interface, and Robot-Robot Interface. Humans have comprehensive real time sensors, e.g., eyes provide visual information and skin provide touching information (contacting force feedback), and fully functional reflection nerves to complete complex tasks in daily life. Therefore, Human-Human Interface is the simplest combination among threes. For Human-Robot Interface, human can finish the action by guiding or following robots` action, the interface has medium difficulty. The most difficult part is Robot-Robot Interface, comprehensive sensors, advanced trajectory planning and tracking, and load distribution algorithms are required for that situation. The more complex the action, the more obvious the effect. The cooperation criterion is denoted by ${C_H}$ for human agent, and ${C_R}$ for manipulator agent.

The cooperation level criterion acts as a coefficient which will be multiplied to agent's safety criterion. The criterion can be determined as following:
\begin{equation}
\label{cooperation_level_equation}
C_P(A_a)=\left\{
\begin{aligned}
(1 + C_H)^{N-1},  if  A_a(a_i) \in A_{human}\\
(1+\frac{C_H + C_R}{2})^{N-1} , \\if A_a(a_i)=\{x|x \in A_{human} \land x \in A_{robot}\}\\
(1 + C_R)^{N-1} , if A_a(a_i) \in A_{Robot}\\
\end{aligned}
\right.
\end{equation}
where $A_a(a_i)$  is the set of agents required to complete certain action $a_i$, and $N$ is the planned number of agents assigned the certain action. 
Integrating all elements determined in this section, the total cost for one agent performing an action can be expressed as:

\begin{equation}
\label{final_cost}
C_{Agent}(a_i) = 1 + F_S + F_I +F_R +R_R + C_I + C_s
\end{equation}

When cooperation between multiple agents are required to finish a certain action, the cost function for a cooperative action will be determined as the average of $C_{Agent}$  based on agent performing the action alone and multiplied by the cooperation criterion as following:

\begin{equation}
\label{final_cost_coop}
C_{Total}(A_a, a_i) = C_P(A_a) \times \frac{1}{N} \sum_{i=1}^{N} C_{Agent}(A_i,a)
\end{equation}

\section{Experiment Results}

\begin{figure}[h!]
\centering
\includegraphics[width=8.5cm]{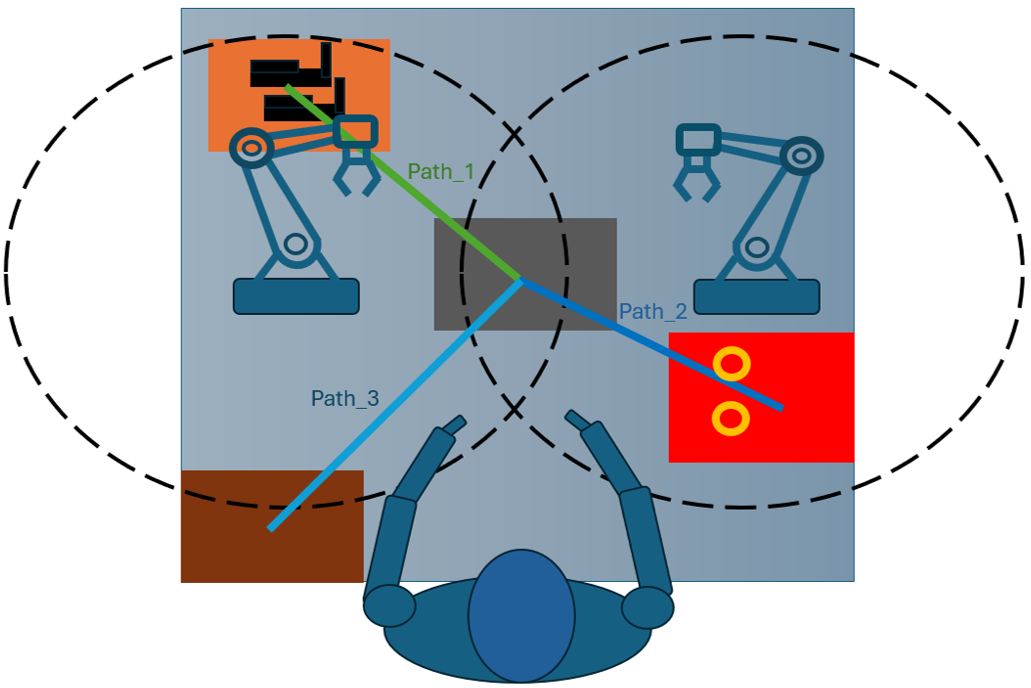}
\caption{Setup Concept Diagram}
\label{Setup Concept include multiple robot manipulators and a human worker}
\end{figure}

The experimental setup concept diagram is displayed in Fig. \ref{Setup Concept include multiple robot manipulators and a human worker}. Two robot manipulators, \textit{robot1} and \textit{robot2} are deployed in the work space and a human worker stands aside of the workspace. The goal is to assemble two parts, whose weight does not exceed the payload capacity of both the human worker and the robot:
\begin{itemize}
    \item Basement parts (135 grams), placed at \textit{storage\_1} (orange area),
    \item Rings (73 grams), placed at the \textit{storage\_2} (red area).
\end{itemize}
The product will be assembled at the workspace (grey area) which is within both robot manipulators' and human worker's range. In the end, assembled products will be moved to \textit{storage\_3} (brown area). 
Paths  between \textit{storage\_1}, \textit{storage\_2}, \textit{storage\_3} and workspace were indicated as \textit{path\_1}, \textit{path\_2}, and \textit{path\_3}. 

\begin{table*}[h!] 

\caption{Costs calculated based on specified experimental setup}
\begin{center}
\label{table_total_cost}
    \begin{tabular}{|c|c|ccc|ccc|ccc|}
    \hline
    & \textbf{Agent} &
    \multicolumn{3}{|c}{Robot\_1} & \multicolumn{3}{|c|}{Robot\_2} & \multicolumn{3}{c|}{Human Worker}\\
    %\hline
     & \textbf{Term} &
    Feasibility & Reachability & Safety &
    Feasibility & Reachability & Safety &
    Feasibility & Reachability & Safety \\
    \hline
    \multirow{4}{*}{PICK/PLACE} & Storage\_1 & \multicolumn{1}{|c}{0+0+0} & \multicolumn{1}{c}{0} & \multicolumn{1}{c}{0} 
                                             & \multicolumn{1}{|c}{0+0+$\infty$} & \multicolumn{1}{c}{N/A} & \multicolumn{1}{c}{N/A} 
                                             & \multicolumn{1}{|c}{0+0+0} & \multicolumn{1}{c}{0} & \multicolumn{1}{c|}{2.976} \\ 
                                & Storage\_2 & \multicolumn{1}{|c}{0+0+$\infty$} & \multicolumn{1}{c}{N/A} & \multicolumn{1}{c}{N/A} 
                                             & \multicolumn{1}{|c}{0+$\infty$/0+0} & \multicolumn{1}{c}{0} & \multicolumn{1}{c}{0} 
                                             & \multicolumn{1}{|c}{0+0+0} & \multicolumn{1}{c}{0} & \multicolumn{1}{c|}{2.976} \\ 
                                & Storage\_3 & \multicolumn{1}{|c}{0+0+0} & \multicolumn{1}{c}{9.09} & \multicolumn{1}{c}{0} 
                                             & \multicolumn{1}{|c}{0+0+$\infty$} & \multicolumn{1}{c}{N/A} & \multicolumn{1}{c}{N/A} 
                                             & \multicolumn{1}{|c}{0+0+0} & \multicolumn{1}{c}{0} & \multicolumn{1}{c|}{0.404} \\ 
                                & Workspace  & \multicolumn{1}{|c}{0+0+0} & \multicolumn{1}{c}{0.455} & \multicolumn{1}{c}{0.637} 
                                             & \multicolumn{1}{|c}{0+0+0} & \multicolumn{1}{c}{0.455} & \multicolumn{1}{c}{0.637} 
                                             & \multicolumn{1}{|c}{0+0+0} & \multicolumn{1}{c}{0} & \multicolumn{1}{c|}{0.637} \\ 
                                \hline
    \multirow{3}{*}{MOVE}       & Path\_1    & \multicolumn{1}{|c}{0+0+0} & \multicolumn{1}{c}{0} & \multicolumn{1}{c}{0.637} 
                                             & \multicolumn{1}{|c}{0+0+$\infty$} & \multicolumn{1}{c}{N/A} & \multicolumn{1}{c}{N/A} 
                                             & \multicolumn{1}{|c}{0+0+0} & \multicolumn{1}{c}{0} & \multicolumn{1}{c|}{2.976} \\
                                & Path\_2    & \multicolumn{1}{|c}{0+0+$\infty$} & \multicolumn{1}{c}{N/A} & \multicolumn{1}{c}{N/A} 
                                             & \multicolumn{1}{|c}{0+0+0} & \multicolumn{1}{c}{0} & \multicolumn{1}{c}{0.637} 
                                             & \multicolumn{1}{|c}{0+0+0} & \multicolumn{1}{c}{0} & \multicolumn{1}{c|}{2.976} \\
                                & Path\_3    & \multicolumn{1}{|c}{0+0+0} & \multicolumn{1}{c}{9.09} & \multicolumn{1}{c}{0} 
                                             & \multicolumn{1}{|c}{0+0+$\infty$} & \multicolumn{1}{c}{N/A} & \multicolumn{1}{c}{N/A}
                                             & \multicolumn{1}{|c}{0+0+0} & \multicolumn{1}{c}{0} & \multicolumn{1}{c|}{0.637} \\
                              
    \hline
    \end{tabular}
\end{center}
\end{table*}

The coordinates of workpieces is predefined, however manipulators don`t know the coordinates of the work-pieces moved by other agents. The action \textbf{PICK} and \textbf{PLACE} are programmed as close and open the gripper, and the action \textbf{MOVE} is programmed as let the end-effector move at a constant speed between two location, however it will lift by 5cm at the beginning of the action to avoid the collision with storage surface. In this experiment, the \textbf{COOPERATE} action is let human provide guidance to \textit{robot2}, which is programmed as reducing the stiffness of \textit{robot2} and let it follow the guidance of human worker. 

\begin{figure}[h!]
\centering
\includegraphics[width=8.5cm]{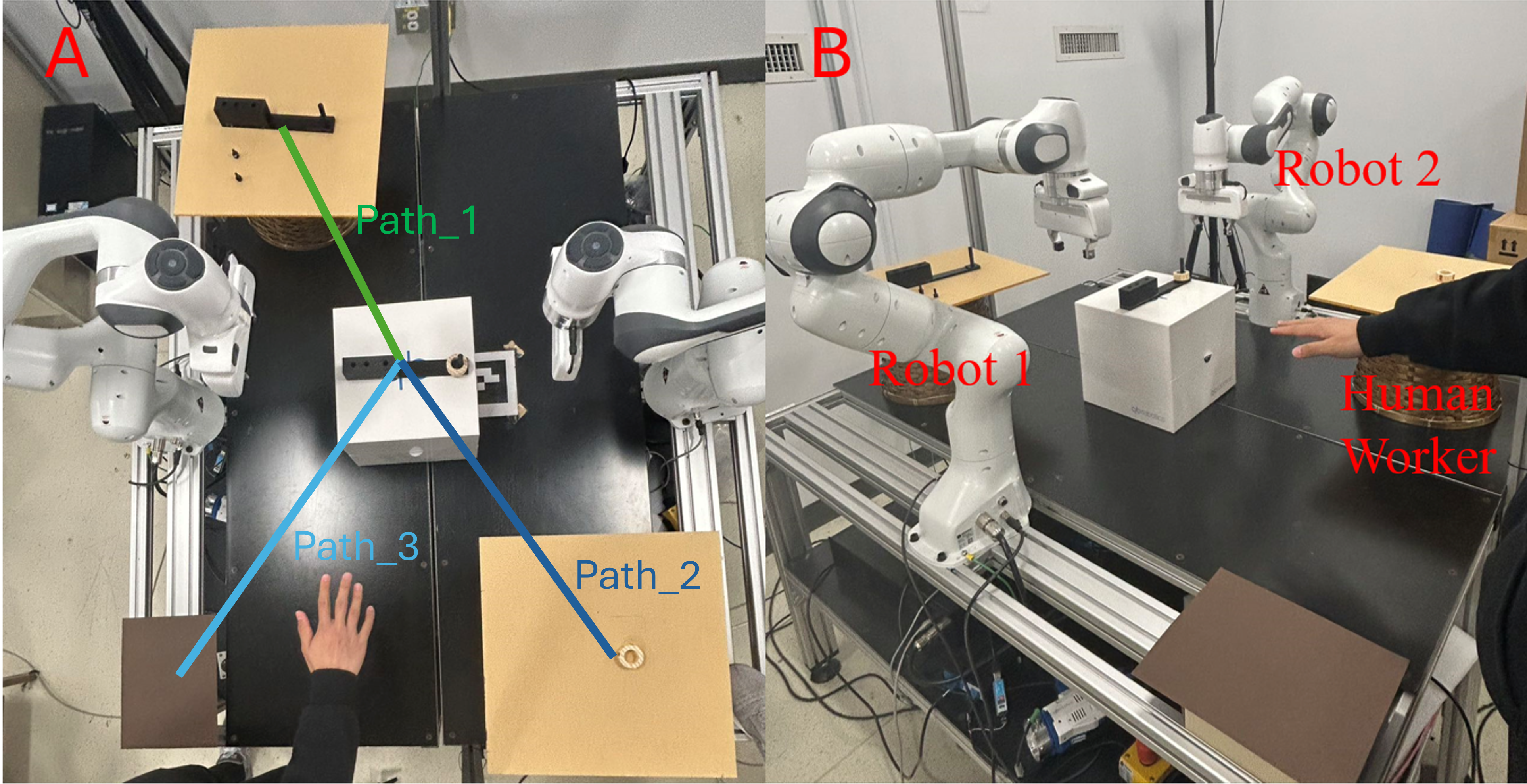}
\caption{Experiment Setup}
\label{Experiment Setup}
\end{figure}

In the experimental setup, the safety control gains ${C_K}$ are specified as follows: for robot agents, ${C_K} = 0.4$ for \textbf{PICK/PLACE}, ${C_K} = 0.6$ for \textbf{MOVE}; for human agents, ${C_K} = 0.1$ for \textbf{PICK/PLACE} and ${C_K} = 0.4$ for \textbf{MOVE}. Move control gain, $K_{C_i}$ is set as a higher value, which is 3, in order to ensure that a human worker only enters the manipulator's workspace when necessary. For reachability cost, the value of reachability index, $D$ is obtained based on  Zacharias’ capability map theory \cite{b29} and expressed in Eq.(\ref{reachability_index}). For \textit{robot1}, \textit{storage\_1} is set in the most suitable work region, where $D > 60$, workspace is set in the suitable region, where $D =55.5$,  \textit{storage\_3} is set in the unsuitable region, where $D = 11.1$.  For \textit{robot2}, \textit{storage\_2} is set in the most suitable work region, where $D > 60$, workspace is set in the suitable region, where $D = 55.5$. The cooperation level criterion are set as ${C_H} = 0.2$ for human workers and ${C_R} = 1$ for robot agents. The Experimental setup is shown in Fig. \ref{Experiment Setup}

Following criterion and equations introduced in Section. \ref{task_allocation}, based on the experimental setup and weights of parts, specific costs for action \textbf{PICK},\textbf{ PLACE}, and \textbf{MOVE} with different parameters are determined and listed as in Table.\ref{table_total_cost}. Feasibility costs listed in Table.\ref{table_total_cost} are in the form of the summation of three feasibility criterions discussed in Section. \ref{agent_strength} to \ref{agent_vlexibility}. For actions with infinity feasibility costs (any one of three criterions), the cost for other terms will not be determined and set to N/A.

As shown in Table. \ref{table_total_cost}, the limitation in robot information level is present when \textit{robot2} deals with \textbf{PICK/PLACE} at workplace. The information level cost is initially infinity since the place coordinate is unknown yet, and become $0$ after human worker guiding the robot to finish the first assembly cycle. Based on Eq. \ref{cooperation_level_equation}, the cooperation level criterion, $C_{P}$,  for the human worker guiding \textit{robot2} is calculated to be 1.6. according to Eq. \ref{final_cost_coop}, the cost of this \textbf{cooperation} action is calculated to be 2.98.

\begin{table}[h!]
\centering
\caption{Total costs of action \textbf{PICK/PLACE} with different parameters }
\label{total_costs_pick}
\begin{tabular}{ccccc}

\hline
%\multicolumn{5}{c}{\textbf{PICK/PLACE}} \\ 
%\hline
\multicolumn{1}{c}{}             & \multicolumn{1}{c}{Storage\_1} & \multicolumn{1}{c}{Storage\_2}              & \multicolumn{1}{c}{Storage\_3} & Workspace \\ 
\hline
\multicolumn{1}{c}{Robot\_1}     & \multicolumn{1}{c}{1}          & \multicolumn{1}{c}{$\infty$}                  & \multicolumn{1}{c}{10.09}      & 2.092     \\ 
%\hline
\multicolumn{1}{c}{Robot\_2}     & \multicolumn{1}{c}{$\infty$}     & \multicolumn{1}{c}{ 1/$\infty$ } & \multicolumn{1}{c}{$\infty$}     & 2.092     \\ 
%\hline
\multicolumn{1}{c}{Human Worker} & \multicolumn{1}{c}{3.976}      & \multicolumn{1}{c}{3.976}                   & \multicolumn{1}{c}{1.404}      & 1.637     \\ 
\hline
\end{tabular}
\end{table}

\begin{table}[h!]
\caption{Total costs of action \textbf{MOVE} with different parameters }
\label{total_costs_move}
\centering
\begin{tabular}{cccc}
\hline
\multicolumn{1}{c}{}             & \multicolumn{1}{c}{Path\_1} & \multicolumn{1}{c}{Path\_2} & \multicolumn{1}{c}{Path\_3} \\%& Workspace \\
\hline
\multicolumn{1}{c}{Robot\_1}     & \multicolumn{1}{c}{1.637}   & \multicolumn{1}{c}{$\infty$}  & \multicolumn{1}{c}{10.09}   \\%& 2.092     \\ 

\multicolumn{1}{c}{Robot\_2}     & \multicolumn{1}{c}{$\infty$}  & \multicolumn{1}{c}{1.637}   & \multicolumn{1}{c}{$\infty$}  \\%& 2.092     \\ 

\multicolumn{1}{c}{Human Worker} & \multicolumn{1}{c}{3.976}   & \multicolumn{1}{c}{3.976}   & \multicolumn{1}{c}{1.637}   \\%& 1.637     \\
\hline
\end{tabular}
\end{table}

Table. \ref{total_costs_pick} - \ref{total_costs_move} contain the total costs determined based on Table. \ref{table_total_cost} and Eq.\ref{final_cost}. Determined total costs will be inputted to task planner as the cost of each specific action.

Two experiments are set up to test and validate of our task allocation algorithm. The first experiment demonstrates the initialization of the production process, where the robot manipulator has not gathered the part coordinates from human. The second experiment demonstrates the first two production cycles, in the second production cycle the robot manipulator has gathered the part coordinates and no longer needs human worker's guidance. Initial and goal states are: 
\begin{itemize}
    \item Base parts and rings are placed at storage as shown in Fig. \ref{Setup Concept include multiple robot manipulators and a human worker},
    \item Robot manipulator \textit{robot1} is initially at \textit{storage\_1}, 
    \item Robot manipulator \textit{robot2} is initially at \textit{workstation},
    \item Human worker \textit{worker} is initially at \textit{storage\_3},
    \item The goal state of the first experiment is to move one assembled part (\textit{finished\_part}) to \textit{storage\_3},
    \item The goal state of the second experiment is to move two assembled parts (\textit{finished\_part}) to \textit{storage\_3}.
\end{itemize}

\begin{figure}[h!]
\centering
\includegraphics[width=9cm]{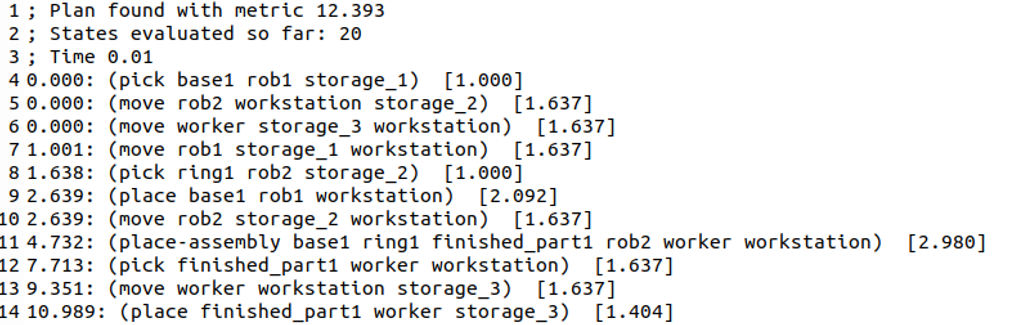}
\caption{Action plan of the first assembly cycle for agents generated by PDDL/POPF solver}
\label{Plan_1}
\end{figure}

\begin{figure}[h!]
\centering
\includegraphics[width=9cm]{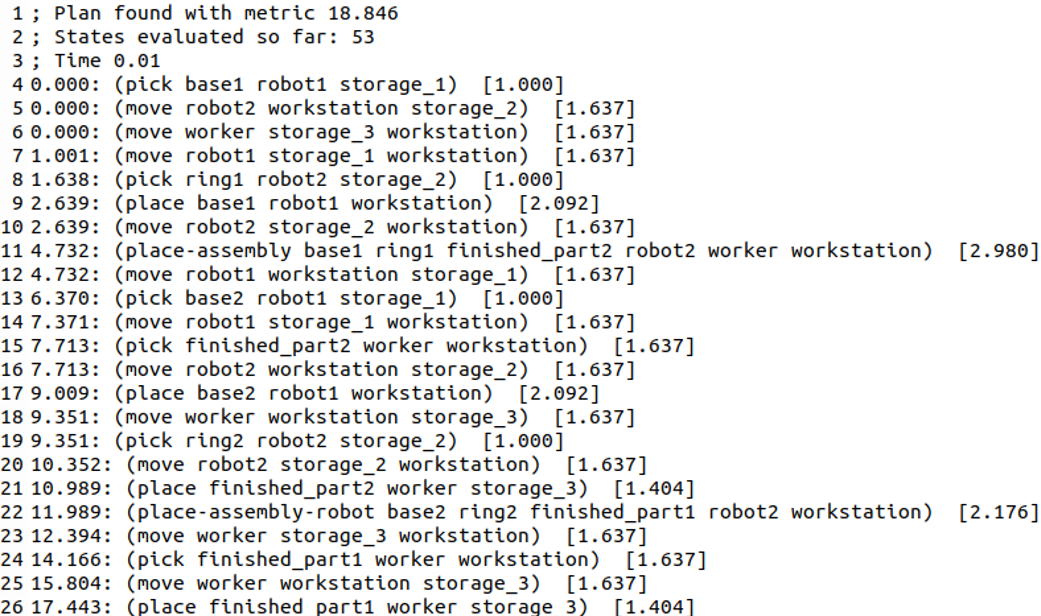}
\caption{Action plan of the first two assembly cycles for agents generated by PDDL/POPF solver}
\label{Plan_2}
\end{figure}

Fig. \ref{Plan_1} shows the optimal plan generated by POPF task planner for the first experiment. Fig. \ref{Plan_2} shows the optimal plan generated for the second experiment. As shown in Line $11$ of Fig. \ref{Plan_2}, the system will request human worker to provide the coordinate of \textit{base\_1} to \textit{robot2} to assemble the ring to the base part after \textit{robot1} has moved it from \textit{storage\_1} to workspace. However, \textit{robot2} has gained the coordinates of the base part in the second assembly cycle, therefore, the system will assign the action with lower cost to \textit{robot2} as shown in Line 22 of Fig. \ref{Plan_2}.

% To use the PDDL/POPF solver to plan the agent actions, a PDDL/POPF domain file is written to simulate the experimental environment. The domain file defines the type of agents, locations, and actions. Costs of actions with different parameters in Table. \ref{total_costs_pick} - \ref{total_costs_move} are programmed into the PDDL/POPF problem file. The PDDL/POPF problem file also defines initial state and goal state of the entire system, which includes followings:

\begin{figure}[h!]
\centering
\includegraphics[width=8.5cm]{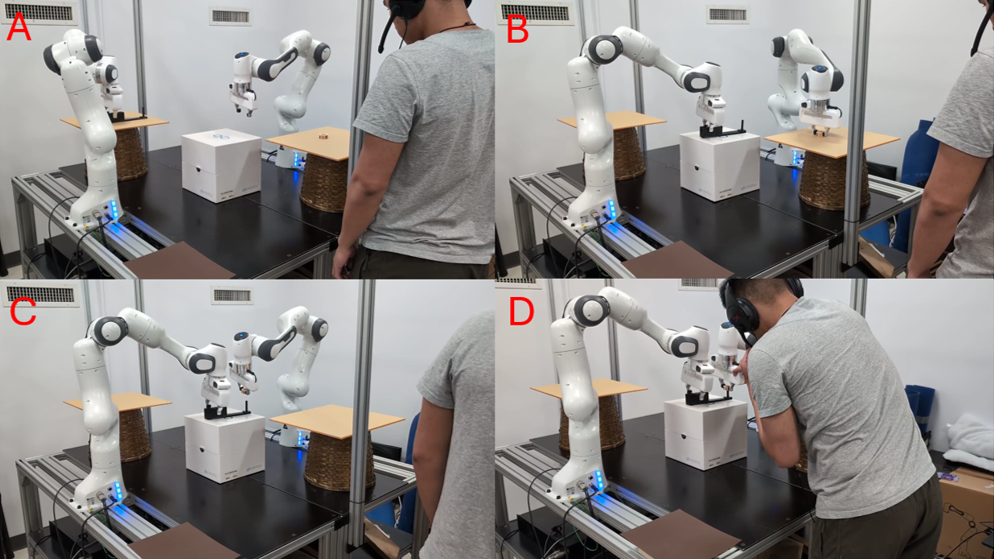}
\caption{Experiment steps: A: the initial state of the system; B: Franka Emika robot manipulator \textit{robot1} moving from \textit{storage\_1} to workstation carrying the base part, (step 4 from Fig. \ref{Plan_1}); C: the robot manipulator \textit{robot2} moving from \textit{storage\_2} to workstation carrying the ring part, (step 7 from Fig. \ref{Plan_1}); D: a human worker guiding manipulator to find the coordinate of the base part on the workstation (step 8 from Fig. \ref{Plan_1}). The Video of the overall experiment is available at: https://youtu.be/P6kif34oJq0}
\label{experiment_process}
\end{figure}

%把fig8和fif6结合起来

The experiment is conducted based on the generated plan to validate and test the task allocation algorithm in the real life situation. Some photos of the experiment are shown in Fig. \ref{experiment_process}. The communication process between the human worker and system is established with Google Speech-to-Text AI \cite{b31}. Human worker will receive an audio order through headphone at the beginning of this action, and human worker will say ``finish" through microphone when he finish the action, and then the system will execute the next step.

\section{Conclusions and Future Work}
A framework for generating optimal plans for collaborative robot manipulators and human workers to replace hard-coded production line plans, integrating the PDDL task planning language, the POPF task planner, and an innovative task allocation algorithm, is proposed in this paper. 
The most important contribution of this study is that once the system is built based on the algorithm introduced, the only required input to the system is an initial state and a goal state, and the system will assign sub-tasks to different agents based on their suitability  to achieve the goal state. No additional complex programming is required for letting the manufacturing system produce different products.

In future, three extensions may be considered based on this study. The first one is to improve task allocation algorithm by integrating the operation time of each action, the position mean square error for manipulator, and the control force feedback to generate a task plan which not only has an optimal logical order and the optimal agent assignment, but also has the shortest work time and the best process performance. 
The second extension could be on implementing the method introduced in this study within environments featuring a diverse range of agents, including scouting drones and mobile manipulators for expansive pick-and-place tasks. %As more agents are incorporated, the  importance of the commander system becomes increasingly  significance.
The third extension may explore and employ various path planning and control algorithms to execute tasks by robot agents. %A method can be developed to select the most suitable planning and control algorithm for each specific action based on different features of employed algorithms.

%\section*{Acknowledgments}
%This should be a simple paragraph before the References to thank those individuals and institutions who have supported your work on this article.

\end{document}